\documentclass[]{aiaa-tc}	

\usepackage{amsmath}
\usepackage{varioref}		
\usepackage{wrapfig}		
\usepackage{threeparttable}	
\usepackage{dcolumn}		
  \newcolumntype{d}{D{.}{.}{-1}}
\usepackage{nomencl}		
  \makenomenclature
\usepackage{subfigure}		
\usepackage{subfigmat}		
\usepackage{fancyvrb}		
  \fvset{fontsize=\footnotesize,xleftmargin=2em}
\usepackage{lettrine}		
\usepackage[dvips]{dropping}	
\usepackage[colorlinks,letterpaper,dvips]{hyperref}
  \hypersetup{citecolor=blue,
              linkcolor=blue,
 	      urlcolor=blue,
	      pdftitle=Joint probability density function modeling of velocity and scalar in turbulence
	               with unstructured grids,
	      pdfauthor=Jozsef Bakosi}

\title{Joint probability density function modeling of velocity and scalar in turbulence
	             with unstructured grids}

\author{
  J.\ Bakosi, P.\ Franzese and Z.\ Boybeyi\\
  {\normalsize\itshape
   George Mason University, Fairfax, VA, 22030, USA}
 }

\AIAApapernumber{YEAR-NUMBER}
\AIAAconference{Conference Name, Date, and Location}
\AIAAcopyright{\AIAAcopyrightD{YEAR}}

\newcommand{\bv}[1]{\mbox{\boldmath$#1$}}
\newcommand{\mean}[1]{\left<#1\right>}

\newcommand*{\Eqr}[1]{(\ref{#1})}
\newcommand*{\Eqre}[1]{Equation~(\ref{#1})}

\newcommand*{\Eqres}[1]{Equations~(\ref{#1})}

\newcommand*{\Fige}[1]{Figure~\ref{#1}}

\begin{document}

\maketitle

\begin{abstract}
In probability density function (PDF) methods a transport equation is solved numerically to compute the time and space dependent probability distribution of several flow variables in a turbulent flow. The joint PDF of the velocity components contains information on all one-point one-time statistics of the turbulent velocity field, including the mean, the Reynolds stresses and higher-order statistics. We developed a series of numerical algorithms to model the joint PDF of turbulent velocity, frequency and scalar compositions for high-Reynolds-number incompressible flows in complex geometries using unstructured grids. Advection, viscous diffusion and chemical reaction appear in closed form in the PDF formulation, thus require no closure hypotheses. The generalized Langevin model (GLM) is combined with an elliptic relaxation technique to represent the non-local effect of walls on the pressure redistribution and anisotropic dissipation of turbulent kinetic energy. The governing system of equations is solved fully in the Lagrangian framework employing a large number of particles representing a finite sample of all fluid particles. Eulerian statistics are extracted at gridpoints of the unstructured mesh. Compared to other particle-in-cell approaches for the PDF equations, this methodology is non-hybrid, thus the computed fields remain fully consistent without requiring any specific treatment. Two testcases demonstrate the applicability of the algorithm: a fully developed turbulent channel flow and the classical cavity flow both with scalars released from concentrated sources.
\end{abstract}

\printnomenclature 

\section{Introduction}
\dropping{2}{P}{\textsc{robability}} density function (PDF) methods \cite{Pope_85,Dopazo_94} belong to the broader family of statistical approaches of turbulence modeling. As opposed to moment closure techniques, in PDF methods the full PDF of the turbulent flow variables is sought, which provides all one-point one-time statistical moments of the underlying fields. Raising the description to higher levels has several advantages. Convection and mean pressure appear in closed form and are treated mathematically exactly. The closure problem can be severe in combustion engineering, where developing accurate closure techniques for the highly nonlinear chemical source terms has proved elusive for realistic configurations. In large eddy simulation (LES), where most of the energy containing motions are sought to be exactly resolved, at high Reynolds number and Damk\"ohler number the chemical reactions take place at subgrid scales. Therefore these processes require closure assumptions in LES as well and the results have a first order dependence on the accuracy of these models \cite{Pope_04}. In the PDF equations the source terms due to chemical reactions appear in closed form, thus no closure assumptions are necessary. Advection and viscous diffusion, processes that are fundamental in near-wall turbulent flows, are also in closed form. Closure hypotheses are necessary for the effect of fluctuating pressure, dissipation of turbulent kinetic energy and small-scale mixing of the scalar. In principle, a more complete statistical description is possible by solving for the full PDF instead of its specific moments as is done in moment closures.

The price to pay for the increased level of description is the high dimensionality of the governing transport equation. As a consequence, instead of employing traditional numerical techniques, such as the finite difference or finite element methods, the Eulerian field equations are written in a Lagrangian form and Monte Carlo methods are used to integrate a set of stochastic differential equations. Numerically, the flow is represented by a large number of Lagrangian particles that represent a finite sample of all fluid particles of the turbulent flow. This methodology not only has the advantage that Monte Carlo techniques are more economical for problems with high dimensionality, but the Lagrangian equations also appear in a significantly simpler form than their Eulerian counterparts.

A natural way of combining existing Eulerian flow solvers with PDF methods is to develop hybrid formulations. Several authors reported on hybrid finite volume (FV)/Monte Carlo algorithms employing both structured \cite{Muradoglu_99,Jenny_01} and unstructured grids \cite{Zhang_04,Rembold_06}. Different types of hybrid algorithms are possible depending on which quantities are computed in the Eulerian and Lagrangian framework and how the information exchange is carried out between the two representations \cite{Muradoglu_99}. However, a common characteristic of these hybrid formulations is that certain consistency conditions have to be met (and enforced on the numerical level) to ensure that all computed fields remain consistent throughout the simulation.

We propose a non-hybrid formulation that is self-consistent both at the level of turbulence closure and the numerical method. Since no fields are computed redundantly, no consistency conditions have to be enforced. The joint PDF of velocity, characteristic turbulent frequency and a set of passive or reactive scalars are computed fully in the Lagrangian framework. An unstructured grid is used \emph{(i)} to extract Eulerian statistics from particles, \emph{(ii)} to solve for inherently Eulerian quantities, such as the mean pressure and \emph{(iii)} to track particles throughout the flow domain. To model the high anisotropy and inhomogeneity of the Reynolds stress tensor in the vicinity of walls Dreeben \& Pope \cite{Dreeben_98} combined Durbin's elliptic relaxation technique \cite{Durbin_93} with the generalized Langevin model (GLM) of Haworth \& Pope \cite{Haworth_86}. We have implemented the model in a general two-dimensional setting for complex geometries. The flow is resolved down to the viscous wall region, by imposing only the no-slip condition on particles without any damping or wall-functions. Two simple testcases are presented: a fully developed turbulent flow in a long-aspect-ratio channel geometry and a turbulent cavity flow, both with scalar releases.

\section{Governing equations}
\label{sec:governing_equations}
\dropping{2}{T}{\textsc{he}} Eulerian governing equation for a viscous incompressible flow is
\begin{equation}
\frac{\partial U_i}{\partial t} + U_j\frac{\partial U_i}{\partial x_j} + \frac{1}{\rho}\frac{\partial P}{\partial x_i} =  \nu\nabla^2U_i \label{eq:NavierStokes},
\end{equation}
where \(U_i\), \(P\), \(\rho\) and \(\nu\) are the Eulerian velocity, pressure, constant density and kinematic viscosity, respectively. The Navier-Stokes equation \Eqr{eq:NavierStokes} is written in the Lagrangian framework as a system of governing equations for Lagrangian particle locations \(\mathcal{X}_i\) and velocities \(\mathcal{U}_i\) \cite{Dreeben_97}
\begin{eqnarray}
\mathrm{d}\mathcal{X}_i&=&\mathcal{U}_i\mathrm{d}t + \left(2\nu\right)^{1/2}\mathrm{d}W_i\label{eq:Lagrangian-position-exact}\\
\mathrm{d}\mathcal{U}_i(t)&=&-\frac{1}{\rho}\frac{\partial P}{\partial x_i}\mathrm{d}t + 2\nu\frac{\partial^2U_i}{\partial x_j\partial x_j}\mathrm{d}t+\left(2\nu\right)^{1/2}\frac{\partial U_i}{\partial x_j}\mathrm{d}W_j,\label{eq:Lagrangian-velocity-exact}
\end{eqnarray}
where the isotropic Wiener process \cite{Gardiner_04} \(\mathrm{d}W_i\), which is a known stochastic process with zero mean and variance \(\mathrm{d}t\), is identical in both equations and the Eulerian fields on the right hand side are evaluated at the particle locations \(\mathcal{X}_i\). The momentum of the particles governed by \Eqres{eq:Lagrangian-position-exact} and \Eqr{eq:Lagrangian-velocity-exact} accounts for both advection and diffusion in physical space with a Gaussian probability distribution. After applying Reynolds decomposition to the Eulerian velocity \(U_i=\mean{U_i}+u_i\) and pressure \(P=\mean{P}+p\) we adopt the generalized Langevin model (GLM) \cite{Haworth_86} to model the appearing unclosed terms and obtain the stochastic model equation for the particle velocity increment
\begin{equation}
\begin{split}
\mathrm{d}\mathcal{U}_i(t)&=-\frac{1}{\rho}\frac{\partial\mean{P}}{\partial x_i}\mathrm{d}t + 2\nu\frac{\partial^2\mean{U_i}}{\partial x_j\partial x_j}\mathrm{d}t + \left(2\nu\right)^{1/2}\frac{\partial\mean{U_i}}{\partial x_j}\mathrm{d}W_j\\
&\quad+G_{ij}\left(\mathcal{U}_j-\mean{U_j}\right)\mathrm{d}t + \left(C_0\varepsilon\right)^{1/2}\mathrm{d}W'_i,
\end{split}
\label{eq:Lagrangian-model}
\end{equation}
where \(G_{ij}\) is a second-order tensor function, \(C_0\) is a positive constant, \(\varepsilon\) denotes the rate of dissipation of turbulent kinetic energy and \(\mathrm{d}W'_i\) is another Wiener process. In general, it is assumed that the tensor \(G_{ij}\) is a function of the Reynolds stress tensor \(\mean{u_iu_j}\), the dissipation rate \(\varepsilon\) and the mean velocity gradients \(\partial\mean{U_i}/\partial x_j\). The last two terms of \Eqre{eq:Lagrangian-model} jointly model the processes of pressure redistribution and dissipation of turbulent kinetic energy. The specification of \(G_{ij}\) determines a particular local closure. Since minimal requirements on \(G_{ij}\) and \(C_0\) automatically guarantee realizablity, \(G_{ij}\) can be specified so that the stochastic equation \Eqr{eq:Lagrangian-model} yields equivalent statistics with any popular Reynolds stress closure at the level of second moments \cite{Pope_94}. For near-wall flows \(G_{ij}\) may also be specified by an elliptic equation based on the analogy that the pressure in incompressible flows is governed by a Poisson equation. This results in a non-locally determined Reynolds stress tensor, where the low-Reynolds-number effects of the wall are felt solely through the boundary conditions of this elliptic equation. Close to the wall, the elliptic operator affects the solution, while far from the wall the solution blends into a local Reynolds stress model. Details on the elliptic relaxation technique are described by Durbin \cite{Durbin_93} in the Reynolds stress framework and by Dreeben \& Pope \cite{Dreeben_98} in the PDF framework.

\Eqre{eq:Lagrangian-model} can be closed by providing length or timescale information for the turbulence. In traditional moment closures a model equation is solved for the turbulent kinetic energy dissipation rate \cite{Hanjalic_72} \(\varepsilon\) or for the mean characteristic turbulent frequency \cite{Wilcox_93} \(\mean{\omega}\) with the definition of the dissipation rate as \(\varepsilon = k\mean{\omega}\), where \(k=\frac{1}{2}\mean{u_iu_i}\) denotes the turbulent kinetic energy. In pure Lagrangian PDF methods, however, an alternative Lagrangian approach has been preferred for the characteristic particle frequency \(\omega\). A model for inhomogeneous flows has been developed by van Slooten \& Pope \cite{vanSlooten_98}, whose simplest form is
\begin{equation}
\mathrm{d}\omega = -C_3\mean{\omega}\left(\omega-\mean{\omega}\right)\mathrm{d}t - S_\omega\mean{\omega}\omega\mathrm{d}t+\left(2C_3C_4\mean{\omega}^2\omega\right)^{1/2}\mathrm{d}W,\label{eq:frequency-model}
\end{equation}
where \(S_\omega\) is a source/sink term for the mean turbulent frequency
\begin{equation}
S_\omega=C_{\omega2}-C_{\omega1}\frac{\mathcal{P}}{\varepsilon},\label{eq:frequency_source}
\end{equation}
where \(\mathcal{P}=-\mean{u_iu_j}\partial\mean{U_i}/\partial x_j\) is the production of turbulent kinetic energy, \(\mathrm{d}W\) is a scalar-valued Wiener-process, while \(C_3,C_4,C_{\omega1}\) and \(C_{\omega2}\) are model constants.

In many practical engineering problems, such as combustion and atmospheric dispersion, the transport and dispersion of passive and reactive scalars in turbulence is of fundamental importance. A remarkable feature of PDF models is that since the source/sink terms in the advection-diffusion equations governing these scalar quantities appear in closed form, they can be represented mathematically exactly, without closure assumptions even in turbulent flows. The Eulerian governing equations for a set of reactive scalars \(\phi_\alpha\), \(\alpha=1\dots n\)
\begin{equation}
\frac{\partial\phi_\alpha}{\partial t} + \bv{U}\cdot\nabla\phi_\alpha = \Gamma\nabla^2\phi_\alpha + S_\alpha(\bv{\phi})
\end{equation}
are written in the Lagrangian framework for the instantaneous particle compositions \(\psi_\alpha\)
\begin{equation}
\mathrm{d}\psi_\alpha = \Gamma\nabla^2\phi_\alpha\mathrm{d}t + S_\alpha(\psi)\mathrm{d}t,
\end{equation}
where the source terms \(S_\alpha(\psi)\) are closed and closure is needed for the molecular diffusion term. For simplicity, the molecular diffusivity \(\Gamma\) is taken to be constant, uniform and the same for each composition. To model the molecular diffusion of the scalars we adopt the interaction by exchange with the conditional mean (IECM) model
\begin{equation}
\mathrm{d}\psi_\alpha = -\frac{1}{t_\mathrm{m}}\left(\psi_\alpha-\mean{\phi_\alpha|\bv{V}}\right)\mathrm{d}t + S_\alpha(\psi)\mathrm{d}t,\label{eq:IECM}
\end{equation}
where \(t_\mathrm{m}\) is a mixromixing timescale, while \(\mean{\phi_\alpha|\bv{V}}=\mean{\phi_\alpha|\bv{U}(\bv{x},t)=\bv{V}}\) denotes the expected value of the mean concentrations conditional on the velocity. For more on the theoretical details, see the reviews compiled by Pope \cite{Pope_85} and Dopazo \cite{Dopazo_94}.

In summary, the flow is represented by a large number of Lagrangian particles; their position \(\mathcal{X}_i\), velocity \(\mathcal{U}_i\), characteristic frequency \(\omega\) and scalar concentrations \(\psi_\alpha\) are governed by \Eqres{eq:Lagrangian-position-exact}, \Eqr{eq:Lagrangian-model}, \Eqr{eq:frequency-model} and \Eqr{eq:IECM}, respectively. These equations are discretized and advanced in time with a forward Euler method. The Eulerian statistics appearing in the equations need to be evaluated at the particle positions. An Eulerian grid discretizes the flow domain and provides the spatial locations where these statistics are estimated. The mean pressure appearing in \Eqre{eq:Lagrangian-model} is computed by a pressure projection algorithm, also utilizing the Eulerian grid.

\section{Two testcases}
\label{sec:results}
\dropping{2}{T}{\textsc{wo}} simple testcases demonstrate the applicability of the method: a fully developed turbulent channel flow and a more complex turbulent cavity flow. In the following two subsections selected statistics of the modeled joint PDF of velocity, frequency and a passive scalar are presented.

\subsection{Channel flow}
\label{sec:channel}
\begin{figure}
\centering
\resizebox{15.2cm}{!}{\input{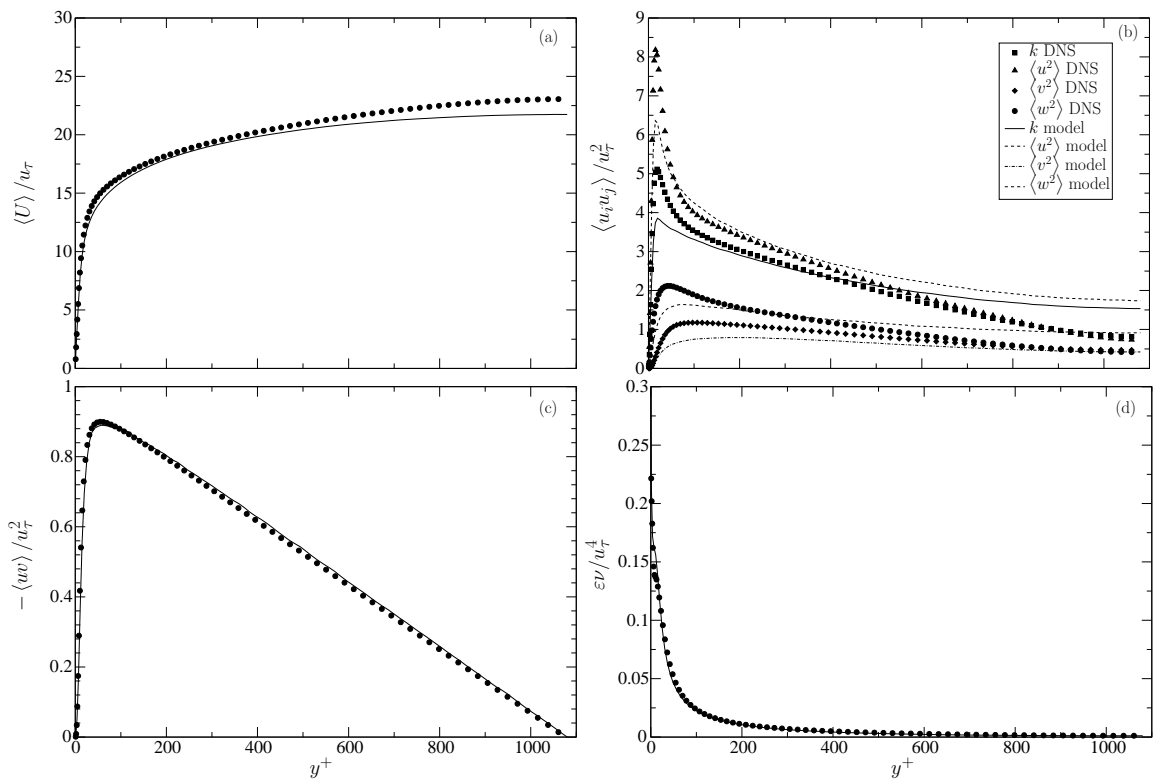}}
\caption{\label{fig:velocity}Cross-stream profiles of (a) the mean streamwise velocity, (b) the diagonal components of the Reynolds stress tensor, (c) the shear Reynolds stress and (d) the rate of dissipation of turbulent kinetic energy. Lines -- PDF calculation, symbols -- DNS data of Abe \emph{et.\ al} \cite{Abe_04}. All quantities are normalized by the friction velocity and the channel half-width. The DNS data is scaled from \(Re_\tau=1020\) to \(1080\).}
\end{figure}

The model has been run for a fully developed turbulent channel flow at the Reynolds number \(Re_\tau=1080\) based on the friction velocity \(u_\tau\) and the channel half width \(h\). After an initial development region this flow becomes statistically stationary, the velocity statistics become one-dimensional and remain inhomogeneous only in the wall-normal direction. Into this flow a passive scalar is released from a concentrated line source located at the channel centerline. Since the scalar field is inhomogeneous, the Eulerian grid is used to compute scalar statistics. Cross-stream profiles of \(\mean{U}\), \(\mean{u_iu_j}\) and \(\varepsilon\) are plotted in \Fige{fig:velocity}. Also shown are the DNS data of Abe \emph{et.\ al} \cite{Abe_04} at \(Re_\tau=1020\).

\begin{figure}
\centering
\resizebox{15.2cm}{!}{\input{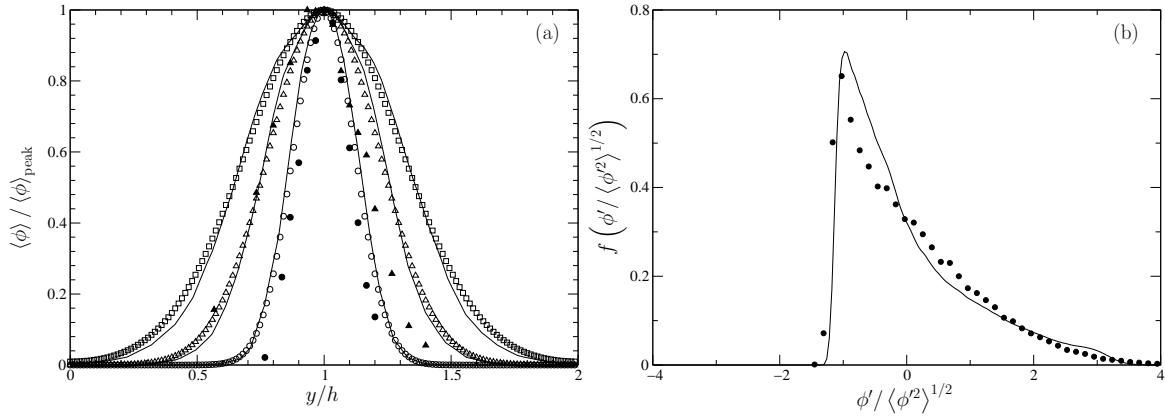}}
\caption{\label{fig:mean-and-pdf}(a) Cross-stream profiles of mean concentration at different downstream locations. Lines -- PDF calculation, hollow symbols -- analytical Gaussians according to Taylor's theory of dispersion \cite{Taylor_21}, filled symbols -- experimental data of Lavertu \& Midlarsky \cite{Lavertu_05}. (b) PDF of concentration fluctuations at a downstream location at the centreline. Lines -- computation, symbols -- experimental data.
}
\end{figure}

In turbulent channel flow the center region of the channel can be considered approximately homogeneous \cite{Brethouwer_01}. Thus for a scalar released at the centreline, Taylor's theory of absolute dispersion \cite{Taylor_21} is expected to describe the mean field of the passive scalar well up to a certain downstream distance from the source. This is shown in \Fige{fig:mean-and-pdf}, where cross-stream mean concentration profiles at different downstream locations are depicted. Also shown in \Fige{fig:mean-and-pdf} is a PDF of scalar concentration fluctuations \(\phi'=\psi-\mean{\phi}\) at a location downstream of the source. The model for the joint PDF of \(\bv{U}\), \(\omega\) and \(\phi\) accurately represents the full PDF and its statistics.

\subsection{Cavity flow}
\label{sec:cavity}
\begin{figure}
\centering
\resizebox{15.2cm}{!}{
\includegraphics{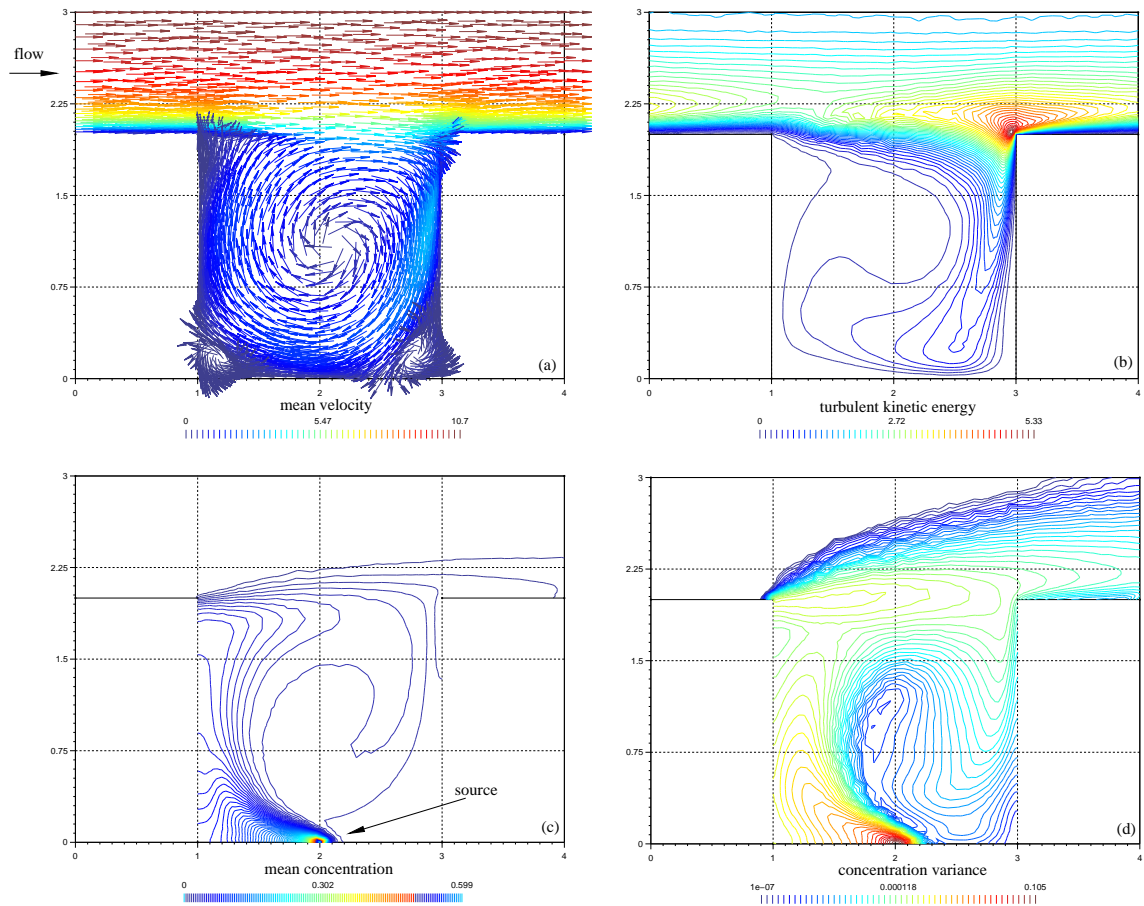}}
\caption{\label{fig:cavity}(a) Geometry and mean velocity distribution for the turbulent cavity flow, (b) spatial distribution of turbulent kinetic energy, (c) mean and (d) variance of scalar concentration. Note the logarithmic scale for the variance. All quantities are normalized by the friction velocity, kinematic viscosity and the concentration at the source.
}
\end{figure}

The turbulent cavity flow is used to demonstrate the applicability of the method in complex geometries. The model has been run at the Reynolds number \(Re=4000\) based on the free stream velocity and cavity depth. This flow also becomes statistically stationary after an initial period. After the flow is fully developed, a passive scalar is released from a concentrated source at the bottom of the cavity. The geometry of the domain and the spatial distributions of mean velocity, turbulent kinetic energy and the mean and variance of the scalar concentration are depicted in \Fige{fig:cavity}.

\section{Conclusion}
\label{sec:conclusion}
We developed and implemented a series of numerical methods to compute the joint PDF of turbulent velocity, frequency and scalar concentrations for high-Reynolds-number incompressible turbulent flows with complex geometries. Adequate wall-treatment on the higher-order statistics is achieved with an elliptic relaxation technique without damping or wall functions. The current examples demonstrate the applicability of the algorithm for two dimensional flows.

\bibliography{jbakosi}
\bibliographystyle{aiaa}

\end{document}